\documentclass[traditabstract]{aa} 
\usepackage{txfonts}
\usepackage[pdftex]{graphicx}
\usepackage{natbib}

\begin{document}
\title{The quest for companions to post-common envelope binaries}
\subtitle{I. Searching a sample of stars from the  CSS and SDSS }

\author{
Backhaus, U. \inst{1} \and 
Bauer, S. \inst{2} \and 
Beuermann, K. \inst{3} \and 
Diese, J. \inst{4} \and 
Dreizler, S\inst{3} \and 
Hessman, F.~V. \inst{3} \and 
Husser, T.-O. \inst{3} \and
Klapdohr, K.-H. \inst{2} \and 
M\"ollmanns, J.$^5$\and 
Sch\"unecke, R. \inst{6} \and 
Dette, J. \inst{2} \and 
Dubbert, J. \inst{2} \and 
Miosga, T. \inst{2} \and 
Rochus Vogel, A.~L. \inst{2} \and
Simons, S. \inst{2} \and 
Biriuk, S. \inst{2} \and 
Debrah, M. \inst{2} \and 
Griemens, M. \inst{2} \and
Hahn, A. \inst{2} \and 
M\"oller, T. \inst{2} \and 
Pawlowski, M. \inst{2} \and 
Schweizer, M. \inst{2} \and 
Speck, A.-L. \inst{2} \and 
Zapros, C. \inst{2} \and 
Bollmann, T. \inst{4} \and
Habermann, F.~N. \inst{4} \and 
Haustovich, N. \inst{4} \and 
Lauser, M. \inst{4} \and 
Liebing, F. \inst{4} \and 
Niederstadt, F. \inst{4} \and 
Hoppen, K. \inst{5} \and
Kindermann, D. \inst{5} \and
K\"uppers, F. \inst{5} \and 
Rauch, B. \inst{5} \and 
Althoff, F. \inst{6} \and 
Horstmann, M. \inst{6} \and 
Kellerman, J.~N. \inst{6} \and 
Kietz, R. \inst{6}, 
Nienaber, T. \inst{6} \and
Sauer, M. \inst{6} \and 
Secci, A. \inst{6} \and 
W\"ullner, L. \inst{6}
}

\institute{
Didaktik der Physik, Universit\"at Duisburg-Essen,Universit\"atsstr. 2, 45117 Essen, Germany \and
Leibniz-Gymnasium, Stankeitstr. 22, 45326 Essen, Germany \and
Institut f\"ur Astrophysik, Georg-August-Universit\"at, Fr.-Hund-Platz 1, 37077 G\"ottingen, Germany \and
Max-Planck-Gymnasium, Theaterplatz 10, 37073 G\"ottingen, Germany \and
Don Bosco-Gymnasium, Theodor-Hartz-Stra{\ss}e 15, Essen, Germany \and
Evangelisches Gymnasium, Beckumer Str. 61, 59555 Lippstadt, Germany	
}

\date{Received 16 September 2011; accepted 19 December 2011}

\authorrunning{U. Backhaus et al.} 
\titlerunning{The quest for companions to post-common envelope binaries I}

\abstract{As part of an ongoing collaboration between student groups
  at high schools and professional astronomers, we have searched for
  the presence of circum-binary planets in a \emph{bona-fide} unbiased
  sample of twelve post-common envelope binaries (PCEBs) from the
  Catalina Sky Survey (CSS) and the Sloan Digital Sky Survey
  (SDSS). Although the present ephemerides are significantly more
  accurate than previous ones, we find no clear evidence for orbital
  period variations between 2005 and 2011 or during the 2011
  observing season. The sparse long-term coverage still permits $O-C$
  variations with a period of years and an amplitude of tens of
  seconds, as found in other systems. Our observations provide the
  basis for future inferences about the frequency with which
  planet-sized or brown-dwarf companions have either formed in
  \mbox{these evolved systems or survived the common envelope (CE)
    phase}.}

\maketitle
 

\section{Introduction}

The detection of circum-binary planets orbiting highly evolved close
binary systems has raised many complex questions about the processes by
which these companions are formed.  These binaries consist of a white
dwarf (WD) or a sub-dwarf B star (sdB) in a tight orbit with a
low-mass secondary star. They result from the rapid orbital evolution,
during which the secondary finds itself immersed in the primary's expanding
red giant envelope \citep{taamricker10}.  This common envelope (CE)
phase results in the loss of a significant amount of orbital angular
momentum and a large part of the primary's original mass.  So far
only a handful of post-common envelope binaries (PCEBs) with
circum-binary planets have been found
\citep[e.g.][]{leeetal09,qianetal09,qianetal10,beuermannetal10,beuermannetal11a,beuermannetal11b},
but other PCEBs display similar orbital period variations
\citep[e.g.][]{parsonsetal10,qianetal11}, giving the impression that
there might be an intimate connection between the evolution of close
binaries, the ejection of the CE, and the presence of planets.  The
case of HU~Aqr indicates a far more complex situation, however. 
\citet{qianetal11} interpreted the $O-C$ variations of this accreting
binary as the combined action of two planets moving around the binary,
but the implied orbits were subsequently shown to be highly unstable
\citep{horneretal11} and led \citet{wittenmyeretal11} to question
their existence.  Obviously, a long-term observation program is needed
to clarify the nature of the eclipse time variations of PCEBs.


\begin{table}[th]
\begin{flushleft}
\caption{New mid-eclipse times and epochs from Drake et al. (2010)}
\begin{tabular}{rcccl}
\hline \\[-1ex]
Cycle   & BJD(TT)      & Error    &  O-C                      & Reference\\
        & 2400000+       & (days)   &  (days)                   & \\[0.5ex]
\hline\\[-1ex]
\multicolumn{5}{l}{\textit{SDSS\,J030308.35+005444.1:}} \\
   13443 & 55798.862876 & 0.000013 & \hspace{-2.4mm}$-$0.000004 & This work \\
   13510 & 55807.870189 & 0.000014 & \hspace{-2.4mm}$-$0.000015 & \\
   13533 & 55810.962273 & 0.000012 &                   0.000002 & \\
   13874 & 55856.805526 & 0.000011 &                   0.000010 & \\
   13897 & 55859.897585 & 0.000011 &                   0.000002 & \\
   13926 & 55863.796278 & 0.000010 &                   0.000003 & \\
   13948 & 55866.753894 & 0.000013 & \hspace{-2.4mm}$-$0.000010 & \\[0.5ex]
\multicolumn{5}{l}{\textit{CSS\,40190~~(SDSS\,J083845.86+191416.5):}}  \\
$-$16141 & 53469.720430 & 0.001300 &  \hspace{-2.4mm}$-$0.000427 & Drake et al.\\
       0 & 55569.862966 & 0.000009 &                  0.000004 & This work \\
     375 & 55618.655072 & 0.000011 &  \hspace{-2.4mm}$-$0.000009 &  \\
    2153 & 55849.994786 & 0.000016 &  \hspace{-2.4mm}$-$0.000000 &  \\
    2214 & 55857.931650 & 0.000013 &                    0.000013 &  \\
    2276 & 55865.998586 & 0.000015 &  \hspace{-2.4mm}$-$0.000015 &  \\[0.5ex]
\multicolumn{5}{l}{\textit{CSS\,03170~~(SDSS\,J085746.18+034255.3):}} \\
$-$31709 & 53464.720527 & 0.000900 &                   0.000020 & Drake et al.\\
       0 & 55528.866640 & 0.000026 & \hspace{-2.4mm}$-$0.000015 & This work \\
     755 & 55578.014506 & 0.000030 & \hspace{-2.4mm}$-$0.000036 &  \\
    1841 & 55648.709375 & 0.000011 & \hspace{-2.4mm}$-$0.000008 &  \\
    1932 & 55654.633181 & 0.000009 &                   0.000013 &  \\
    4887 & 55846.993443 & 0.000026 &                   0.000003 &  \\
    4948 & 55850.964311 & 0.000020 & \hspace{-2.4mm}$-$0.000018 &  \\
    5178 & 55865.936535 & 0.000019 &                   0.000002 &  \\[0.5ex] 
\multicolumn{5}{l}{\textit{CSS\,080502~~(SDSS\,J090812.03+060421.2):}} \\
$-$14073 & 53466.830927 & 0.001400 & \hspace{-2.4mm}$-$0.003100 & Drake et al.\\
       0 & 55569.876004 & 0.000010 & \hspace{-2.4mm}$-$0.000008 & This work \\
      47 & 55576.899608 & 0.000013 &                   0.000006 &  \\
      94 & 55583.923204 & 0.000016 &                   0.000013 &  \\
     514 & 55646.687183 & 0.000011 &                   0.000002 &  \\
     521 & 55647.733251 & 0.000005 &                   0.000003 &  \\
     527 & 55648.629867 & 0.000005 & \hspace{-2.4mm}$-$0.000009 &  \\
     561 & 55653.710777 & 0.000012 &                   0.000006 &  \\
     567 & 55654.607405 & 0.000005 &                   0.000006 &  \\
    1968 & 55863.970136 & 0.000006 & \hspace{-2.4mm}$-$0.000001 & \\
    1981 & 55865.912837 & 0.000007 &                   0.000004 & \\
    1995 & 55868.004964 & 0.000005 & \hspace{-2.4mm}$-$0.000002 & \\[0.5ex]
\multicolumn{5}{l}{\textit{CSS\,38094~~(SDSS\,J093947.95+325807.3):}} \\
 $-$6320 & 53495.954128 & 0.003300 & \hspace{-2.4mm}$-$0.000078 & Drake et al.\\
       0 & 55587.808817 & 0.000035 & \hspace{-2.4mm}$-$0.000006 & This work \\
      84 & 55615.611957 & 0.000010 &                   0.000003 & \\
     202 & 55654.668723 & 0.000021 & \hspace{-2.4mm}$-$0.000011 & \\
     774 & 55843.994813 & 0.000019 & \hspace{-2.4mm}$-$0.000003 & \\
     795 & 55850.945601 & 0.000013 &                   0.000002 & \\[0.5ex]
\multicolumn{5}{l}{\textit{CSS\,41631~~(SDSS\,J095719.24+234240.7):}} \\
$-$14145 & 53470.764125 & 0.001500 &                   0.000612 & Drake et al.\\
       0 & 55604.830127 & 0.000029 &                   0.000003 & This work \\
      72 & 55615.692822 & 0.000008 &                   0.000005 &  \\[1.3ex]
\hline\\[-5ex]
\end{tabular}
\label{tab:data1}
\end{flushleft}
\end{table}

\begin{table}[thb]
\begin{flushleft}

\vspace{8.0mm}
\begin{tabular}{rcccl}
\hline \\[-1ex]
Cycle   & BJD(TT)      & Error    &  O-C                      & Reference\\
        & 2400000+        & (days)   &  (days)                   & \\[0.6ex]
\hline\\[-1ex]
\multicolumn{5}{l}{\textit{CSS\,41631  continued:}} \\
     325 & 55653.863107 & 0.000009 & \hspace{-2.4mm}$-$0.000007 &  \\
     370 & 55660.652293 & 0.000015 & \hspace{-2.4mm}$-$0.000005 &  \\
    1678 & 55857.991223 & 0.000009 & \hspace{-2.4mm}$-$0.000002 &  \\
    1691 & 55859.952549 & 0.000010 &                   0.000005 &  \\[0.5ex]
\multicolumn{5}{l}{\textit{CSS\,41177~~(SDSS\,J100559.10+224932.2):}}\\
$-$18521 & 53470.704027 & 0.001700 & \hspace{-2.4mm}$-$0.000528 & Drake et al.\\
     116 & 55632.884239 & 0.000014 &                   0.000003 & This work \\
     252 & 55648.662338 & 0.000009 &                   0.000003 &  \\
     355 & 55660.611914 & 0.000019 & \hspace{-2.4mm}$-$0.000011 &  \\
    1996 & 55850.993250 & 0.000021 & \hspace{-2.4mm}$-$0.000005 &\\
    2125 & 55865.959247 & 0.000010 & \hspace{-2.4mm}$-$0.000000 & \\[0.5ex]
\multicolumn{5}{l}{\textit{CSS\,21616~~(SDSS\,J132518.18+233808.0):}} \\
$-$11198 & 53470.804327 & 0.001900 & \hspace{-2.4mm}$-$0.000054 & Drake et al.\\
       0 & 55653.954195 & 0.000011 &                   0.000008 & This work \\
     138 & 55680.858500 & 0.000014 & \hspace{-2.4mm}$-$0.000017 &  \\
     215 & 55695.870287 & 0.000056 & \hspace{-2.4mm}$-$0.000067 &  \\
     783 & 55806.607031 & 0.000031 &                   0.000014 &\\[0.5ex]
\multicolumn{5}{l}{\textit{CSS\,06653~~(SDSS\,J132925.21+123025.4):}} \\
$-$26458 & 53466.817520 & 0.000400 &                   0.000499 & Drake et al.\\
       0 & 55609.022163 & 0.000005 & \hspace{-2.4mm}$-$0.000002 & This work \\
      37 & 55612.017920 & 0.000009 &                   0.000003 &  \\
    1034 & 55692.741265 & 0.000006 & \hspace{-2.4mm}$-$0.000008 &  \\
    1059 & 55694.765440 & 0.000007 &                   0.000011 &  \\
    1060 & 55694.846403 & 0.000012 &                   0.000008 &  \\
    2391 & 55802.612475 & 0.000010 & \hspace{-2.4mm}$-$0.000004 & \\[0.5ex]
\multicolumn{5}{l}{\textit{WD1333+005~~(SDSS\,J133616.05+001732.6):}} \\
$-$17605 & 53464.891227 & 0.000600 &  \hspace{-2.4mm}$-$0.001316 & Drake et al.\\
       0 & 55611.976665 & 0.000010 &  \hspace{-2.4mm}$-$0.000002 & This work \\
     679 & 55694.786674 & 0.000010 &                    0.000003 &  \\
    1563 & 55802.598220 & 0.000013 &  \hspace{-2.4mm}$-$0.000002 & \\[0.5ex]
\multicolumn{5}{l}{C\textit{CSS\,06833~~(SDSS\,J153349.44+375928.0):}} \\
$-$13173 & 53480.923730 & 0.000800 & \hspace{-2.4mm}$-$0.000671 & Drake et al.\\
       0 & 55611.926569 & 0.000010 & \hspace{-2.4mm}$-$0.000011 & This work \\
     260 & 55653.986899 & 0.000009 &                   0.000002 &  \\
     439 & 55682.943820 & 0.000009 &                   0.000011 &  \\
     457 & 55685.855660 & 0.000009 & \hspace{-2.4mm}$-$0.000017 &  \\
     488 & 55690.870566 & 0.000009 &                   0.000006 &  \\
    1204 & 55806.698204 & 0.000010 & \hspace{-2.4mm}$-$0.000001 & \\
    1537 & 55860.567775 & 0.000012 &                   0.000009 & \\[0.5ex]
\multicolumn{5}{l}{\textit{SDSS~J154846.00+405728.7:}} \\
$-$5745 & 53526.785031 & 0.002700 & \hspace{-2.4mm}$-$0.002547 & Drake et al.\\   
   5920 & 55690.823510 & 0.000012 &                 0.000013 & This work \\     
   5942 & 55694.904830 & 0.000010 & \hspace{-2.4mm}$-$0.000003 &  \\              
   5947 & 55695.832395 & 0.000016 & \hspace{-2.4mm}$-$0.000015 &  \\
   6507 & 55799.720977 & 0.000023 &                   0.000002 & \\
   6512 & 55800.648543 & 0.000011 & \hspace{-2.4mm}$-$0.000009 & \\
   6523 & 55802.689237 & 0.000015 &                   0.000017 & \\[0.8ex]
\hline \\[-5ex]
\end{tabular}
\label{tab:data2}
\end{flushleft}
\end{table}

{\it PlanetFinders} is a research project conducted by high school
teachers and their students in collaboration with professional
astronomers.  The scientific goal is the measurement of accurate
ephemerides of eclipsing PCEBs with the aim of detecting circum-binary
companions by the light travel time effect. The didactic goal is to
let high school students experience all aspects of authentic
scientific work at an early age.

For the first observing season, we chose to survey twelve eclipsing
PCEBs, eleven of which are from the list of \citet{drakeetal10} drawn
from the Catalina Sky Survey (CSS) and the Sloan Digital Sky Survey
(SDSS). Our PCEBs include ten with a WD primary, CSS\,06833 with an
sdB primary, and the double degenerate CSS\,41177
\citep{parsonsetal11}. All these stars have well-established orbital
periods, mostly obtained in the 2005 observing season by
\citet{drakeetal10}. So far, all of them lack known or suspected
period variations.  Hence, in this respect, the sample is
\emph{bona-fide} unbiased. Obtaining an independent measurement of the
binary period in a single second observing season precludes the ready
measurement of the orbital period of a putative circum-binary
companion, but allows the detection of a period variation. Ultimately,
additional eclipse-time measurements will lead to the discovery of any
companion that creates a sufficiently large period variation.

\section{Observations and data analysis}

All data presented here were taken with the remotely controlled 1.2-m
MONET/North telescope at the University of Texas' McDonald Observatory
via the MONET browser-based remote-observing interface.  The teachers
and their students usually observed from the classrooms of the
participating schools.  The photometric data presented here were taken
with an Apogee ALTA E47+ 1k$\times$1k CCD camera in white light.  Exposure
times were typically 10\,s, separated by a 3-s readout for data binned
in 2$\times$2 pixels, but in some cases exposure times of 15 or 20\,s were
chosen. The images were corrected for dark current and flatfielded in
the usual manner.
 
\begin{figure*}[t]
\includegraphics[bb=140 20 510 700,height=86mm,angle=270,clip]{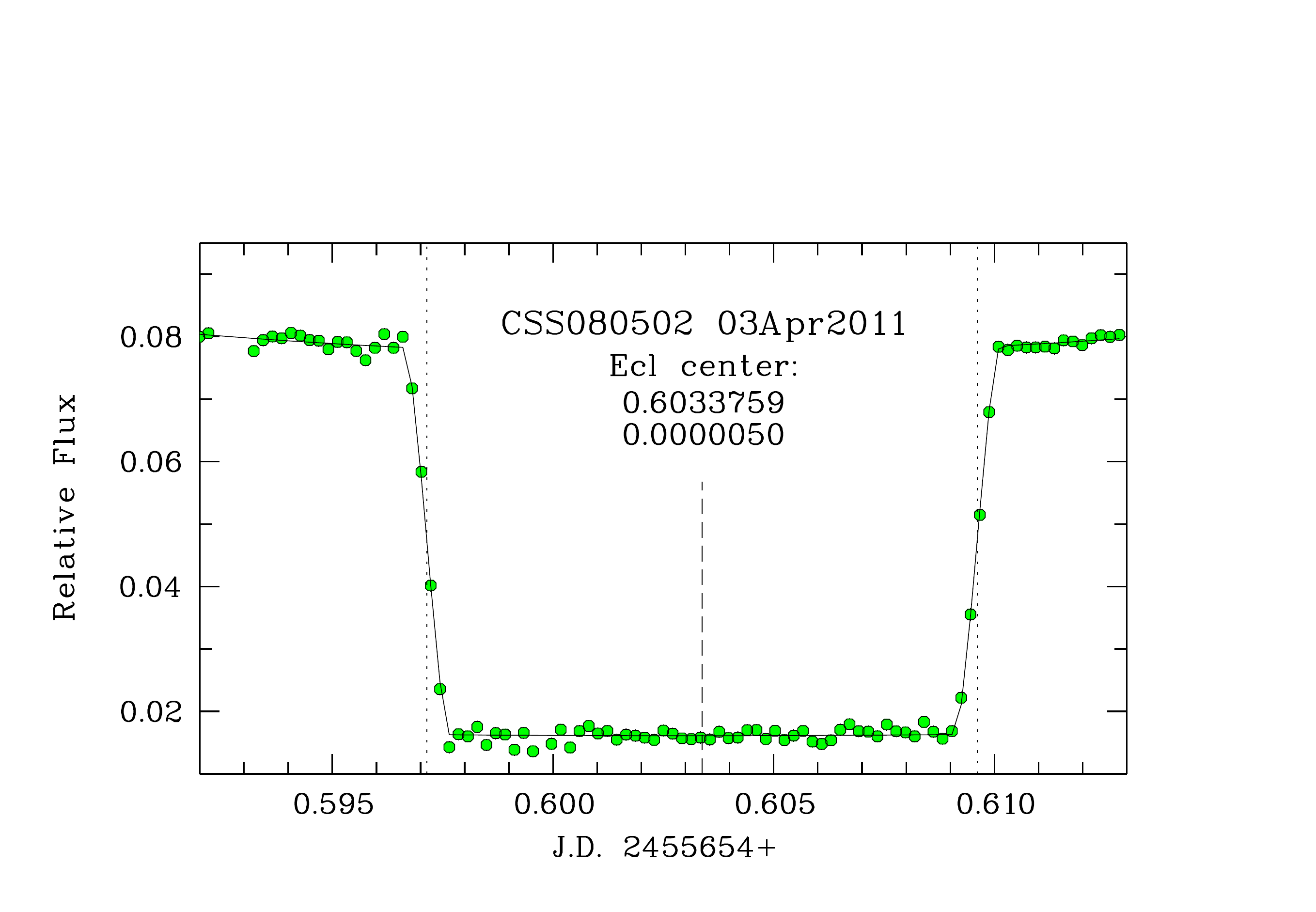}
\hfill
\includegraphics[bb=140 20 510 700,height=86mm,angle=270,clip]{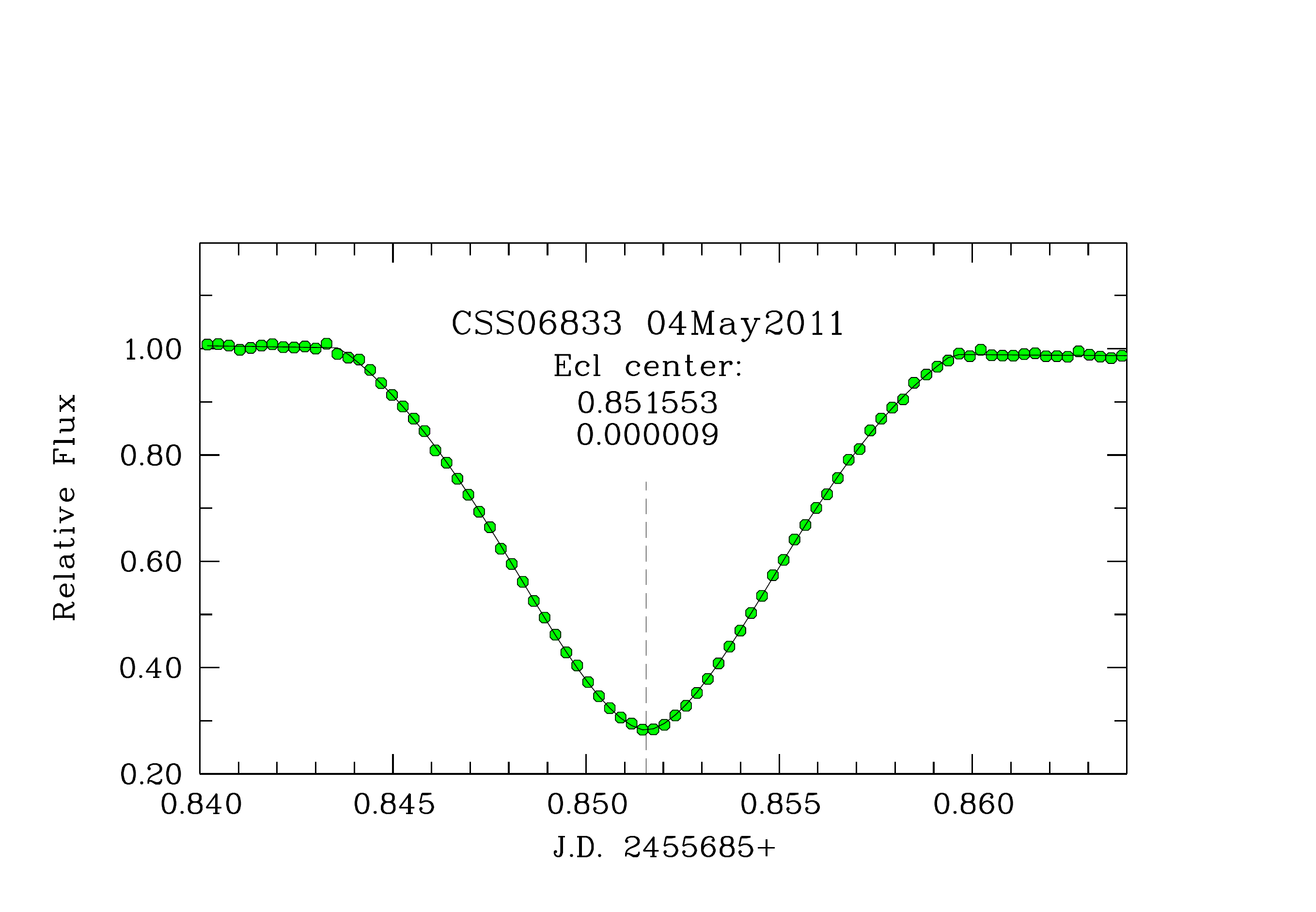}
\caption[chart]{Eclipse light curves of the WD/dM binary CSS\,080502
  (SDSS\,J0908+0604) and the sdB/dM binary CSS\,06833
  (SDSS\,J1533+3759). Exposure times are 15\,s. The solid lines
  represent the model fits, the dotted lines in the left panel
  indicate the FWHM.}
\label{fig:lc}
\end{figure*}

The WD binaries show well-defined eclipse light curves with ingress
and egress times of about a minute.  Usually, we determined the flux
of the target relative to an appropriate non-variable comparison star.
The high-school students employed a variety of methods for determining
the mid-eclipse times $T_\mathrm{ecl}$ for a given light curve, from
graphical ones in the 10th grade to formal fits using a series of
concatenated linear functions in the 12th grade.  For publication, all
data were subjected to more formal fits, which took account of the
finite exposure times and yielded formal errors for the mid-eclipse
times. The adopted models assume symmetry of the eclipse light curve
about mid-eclipse, with ingress and egress taken to be mirror images
of each other. In addition, we allowed for a time-dependent
multiplicative factor that describes a real or apparent variation of
the out-of-eclipse flux by a first- or second-order polynomial. A real
effect arises, e.g., from the varying aspect of the illuminated
secondary star, an apparent effect from the color-dependent and
altitude-dependent atmospheric transmission.  For the binaries with WD
primary, the eclipsed star was represented by a uniform disk with the
ingress/egress time being one of the free parameters.  The eclipse
light curves of the double degenerate system CSS\,41177 and the sdB/dM
binary CSS\,06833 were fitted by a heuristic model that involves a
modified and truncated inverted Gaussian \citep[see][paper
  II]{beuermannetal11b}.  The Gaussian was modified by replacing the
square in the exponential with a free parameter $p_\mathrm{exp}$,
allowing the creation of a more peaked ($p_\mathrm{exp}\!<\!2$) or
broader ($p_\mathrm{exp}\!>\!2$) light curve, and then truncated at
the out-of-eclipse level (one of the other fit parameters).  The
fits to the light curves of CSS\,06833 yield
$p_\mathrm{exp}\!\simeq\!1.70$.  Fig.~\ref{fig:lc} shows examples of the observed
and fitted light curves.

\section{Results}

Between November 2010 and October 2011, a total of 72 eclipse
light curves of our twelve targets were secured. The new mid-eclipse
times are listed in Table~1 as BJD(TT), i.e., barycentrically and
leap-second corrected times in the terrestrial
system\footnote{http://astroutils.astronomy.ohio-state.edu/time/}. The
mean statistical 1-$\sigma$ error of our mid-eclipse times is 1.4\,s,
with individual values ranging from 0.4 to 4.8\,s. The individual
eclipse times obtained by \citet{drakeetal10} in 2005 have not been
published, but we list their epochs converted from HJD to BJD(TT) as
representative eclipse times in Table~1. In deriving new ephemerides,
we did not include the Drake et al. epochs. Hence, our results are
completely independent of theirs. However, we did include additional
published mid-eclipse times for CSS\,41177 \citep{parsonsetal11},
CSS\,06833 \citep{foretal10}, SDSS\,J0303+0054 \citep{parsonsetal10},
SDSS\,J1548+4057 \citep{pyrzasetal09}, and SDSS\,J0303+0054
\citep{parsonsetal10}. In the case of SDSS\,J0303+0054, we used the
accurate ULTRACAM data of \citet{parsonsetal10}, but not the less
accurate timings of Pyrzas et al., which have almost no influence on
the derived ephemeris. The new epochs and periods are given in Table~2
along with those of \citet{drakeetal10}. The accuracy of the new
periods exceeds those of \citet{drakeetal10} by one to two orders of
magnitude. The \,$\chi^2$ of the linear fit and the number of degrees
of freedom (d.o.f.) are given in the last column.

\begin{figure*}[t]
\includegraphics[bb=260 28 465 700,height=64mm,width=20mm,angle=-90,clip]{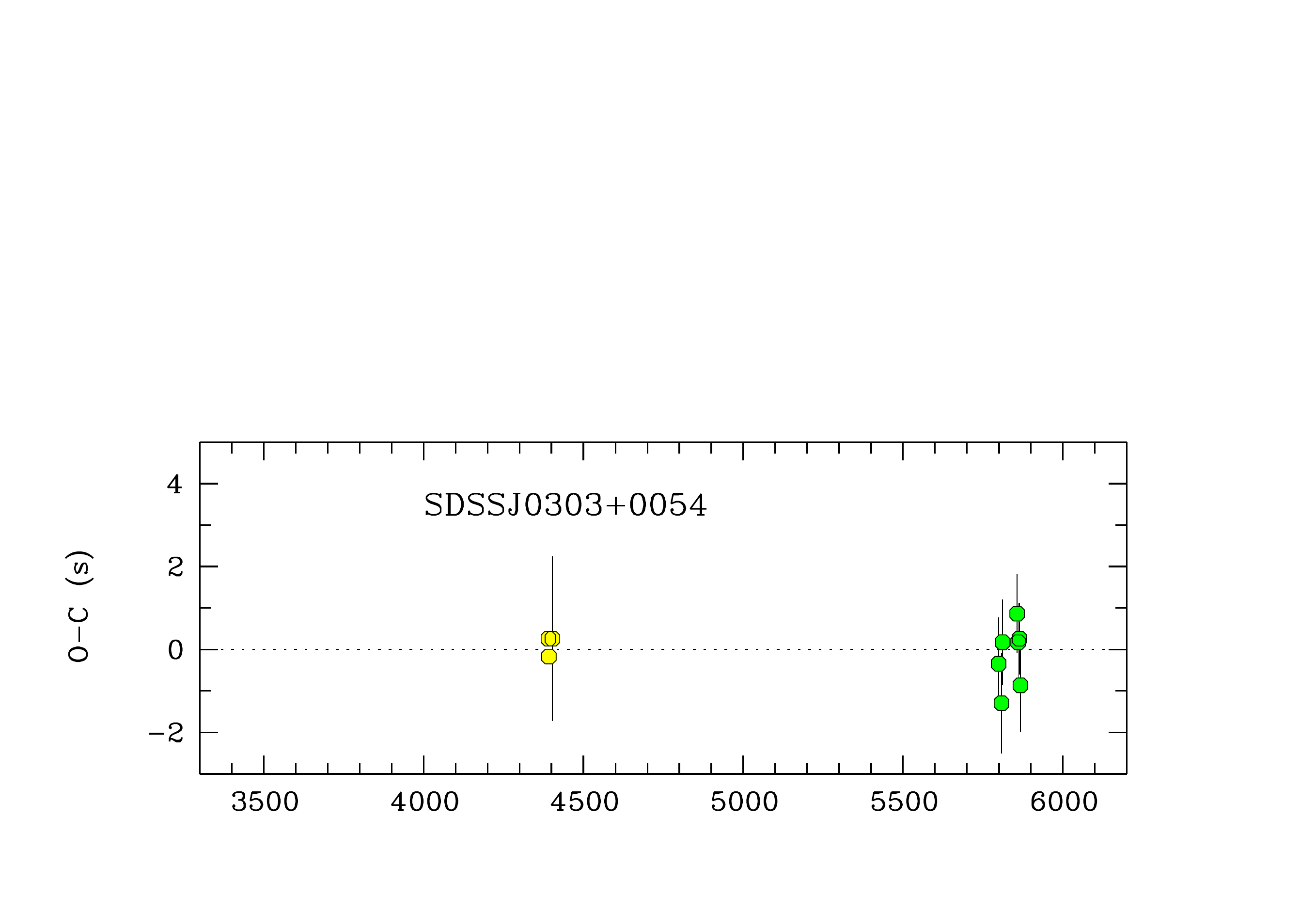}
\includegraphics[bb=260 79 465 700,height=60mm,width=20mm,angle=-90,clip]{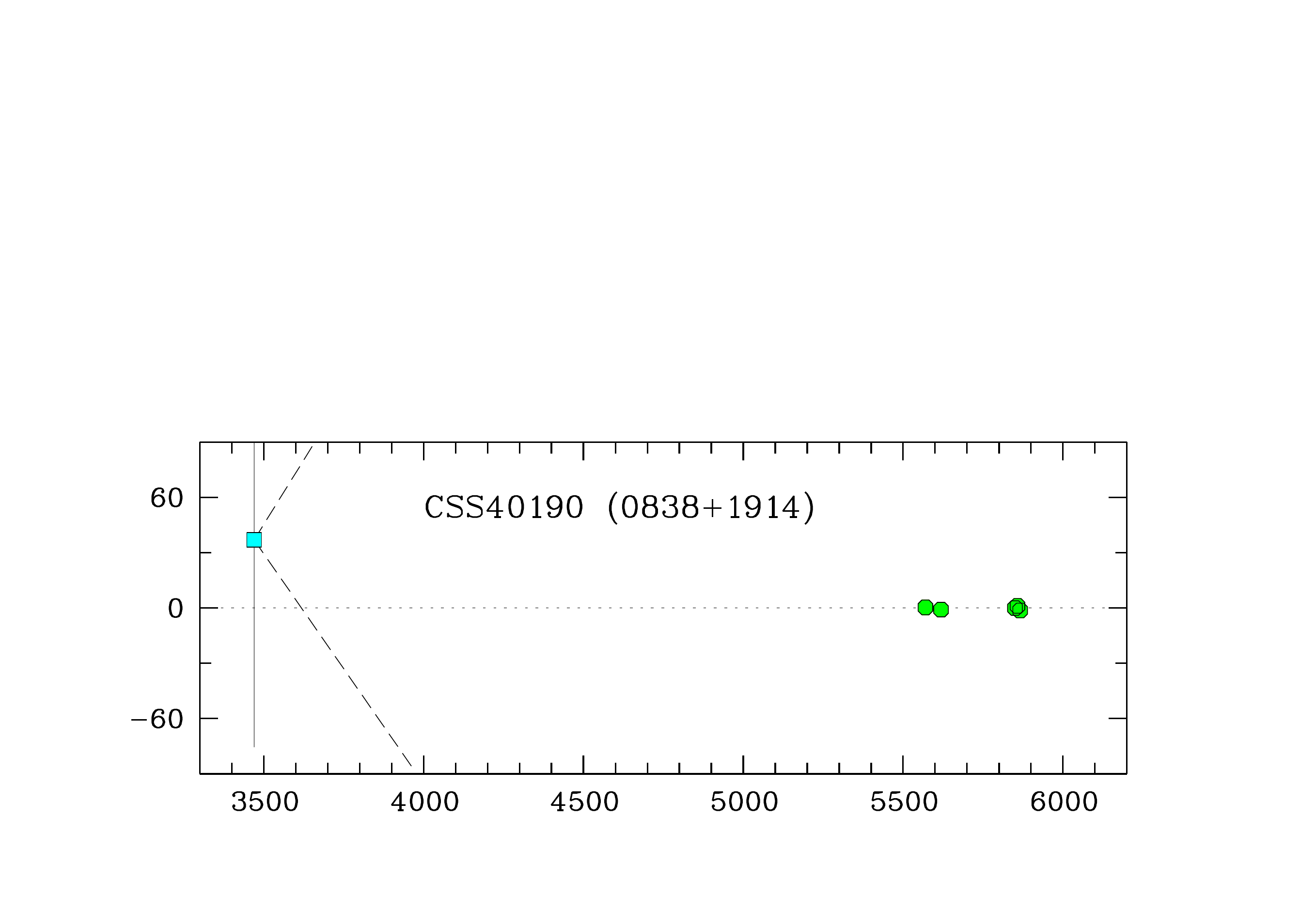}
\includegraphics[bb=260 79 465 700,height=60mm,width=20mm,angle=-90,clip]{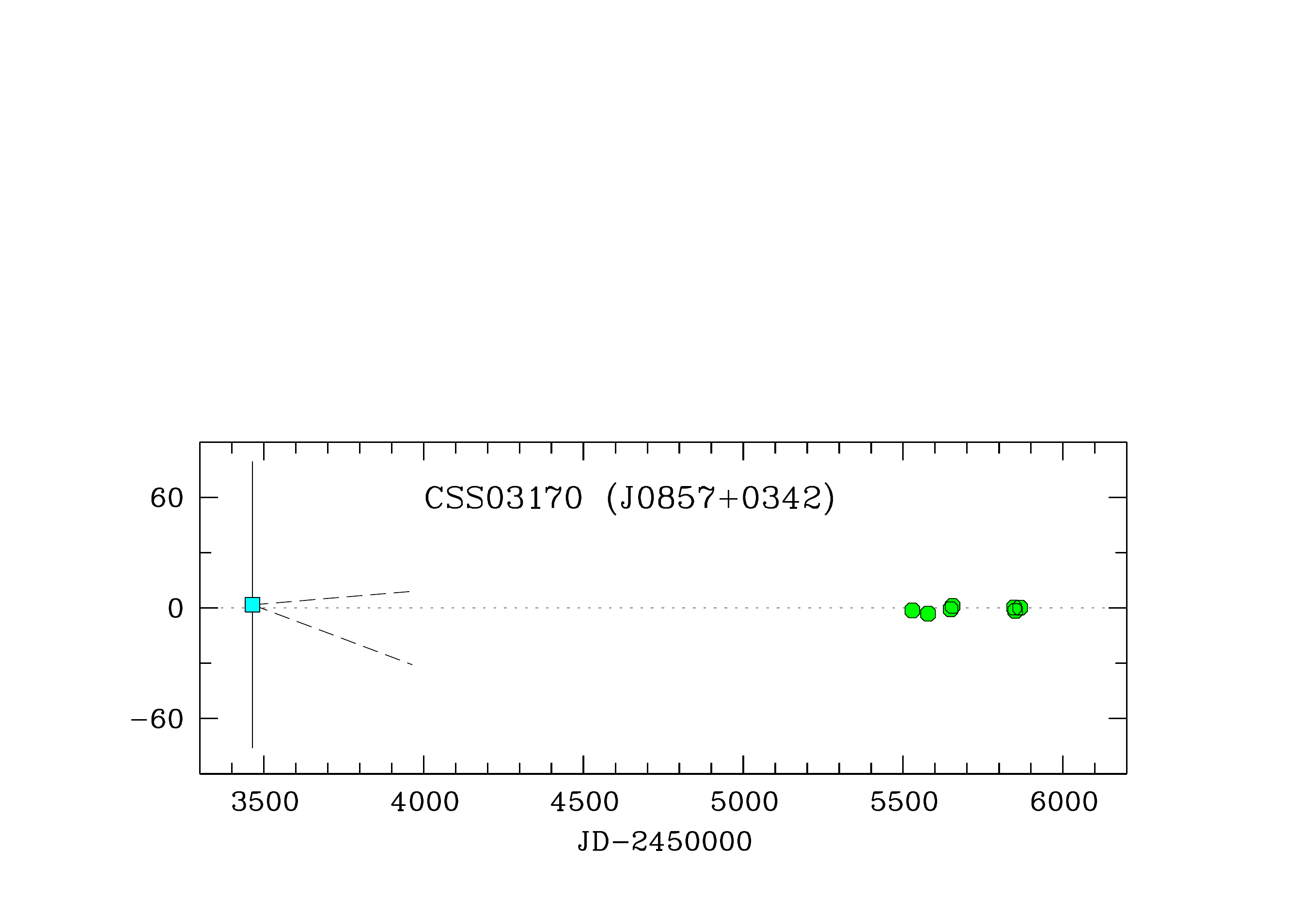}

\includegraphics[bb=260 28 465 700,height=64mm,width=20mm,angle=-90,clip]{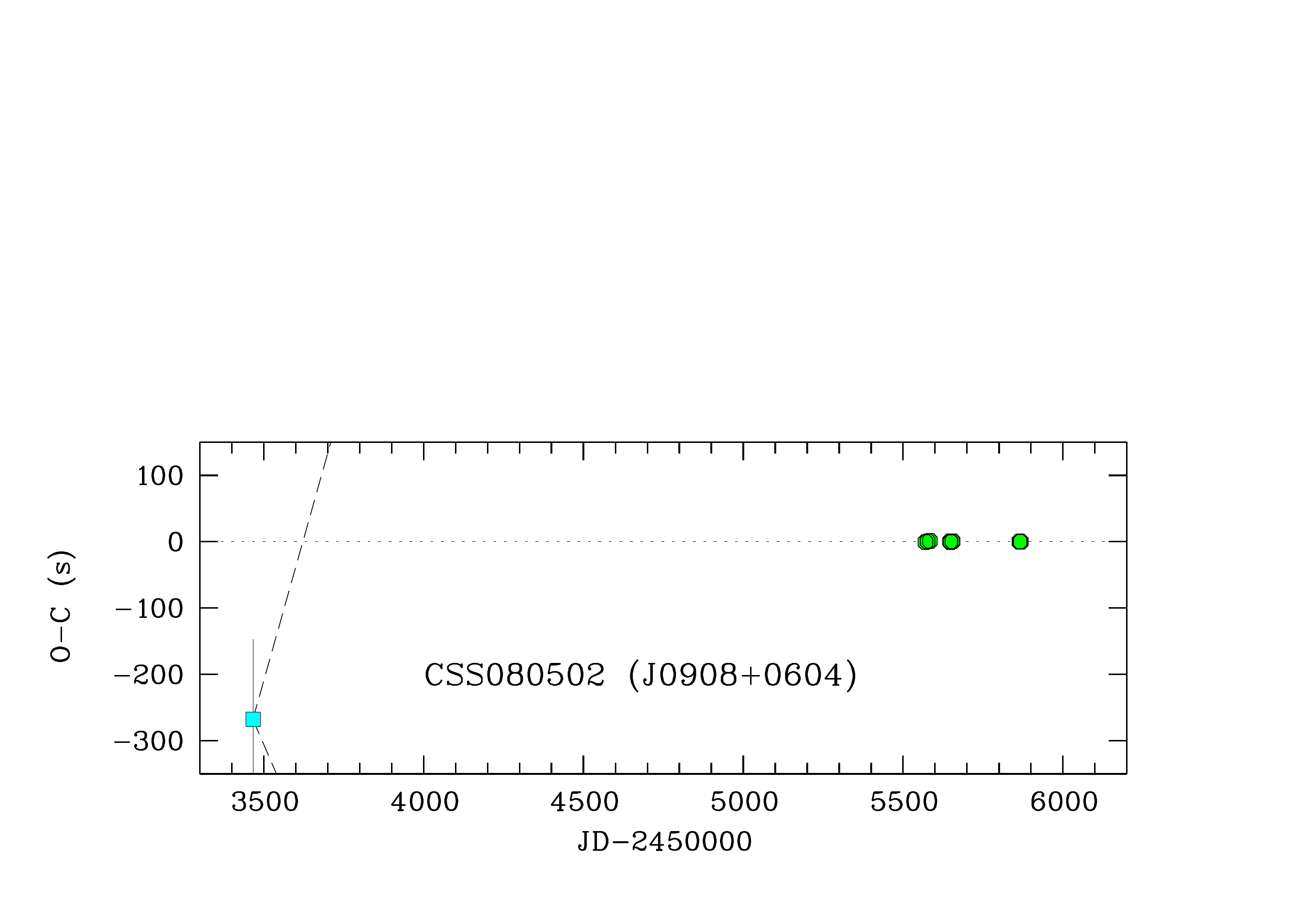}
\includegraphics[bb=260 79 465 700,height=60mm,width=20mm,angle=-90,clip]{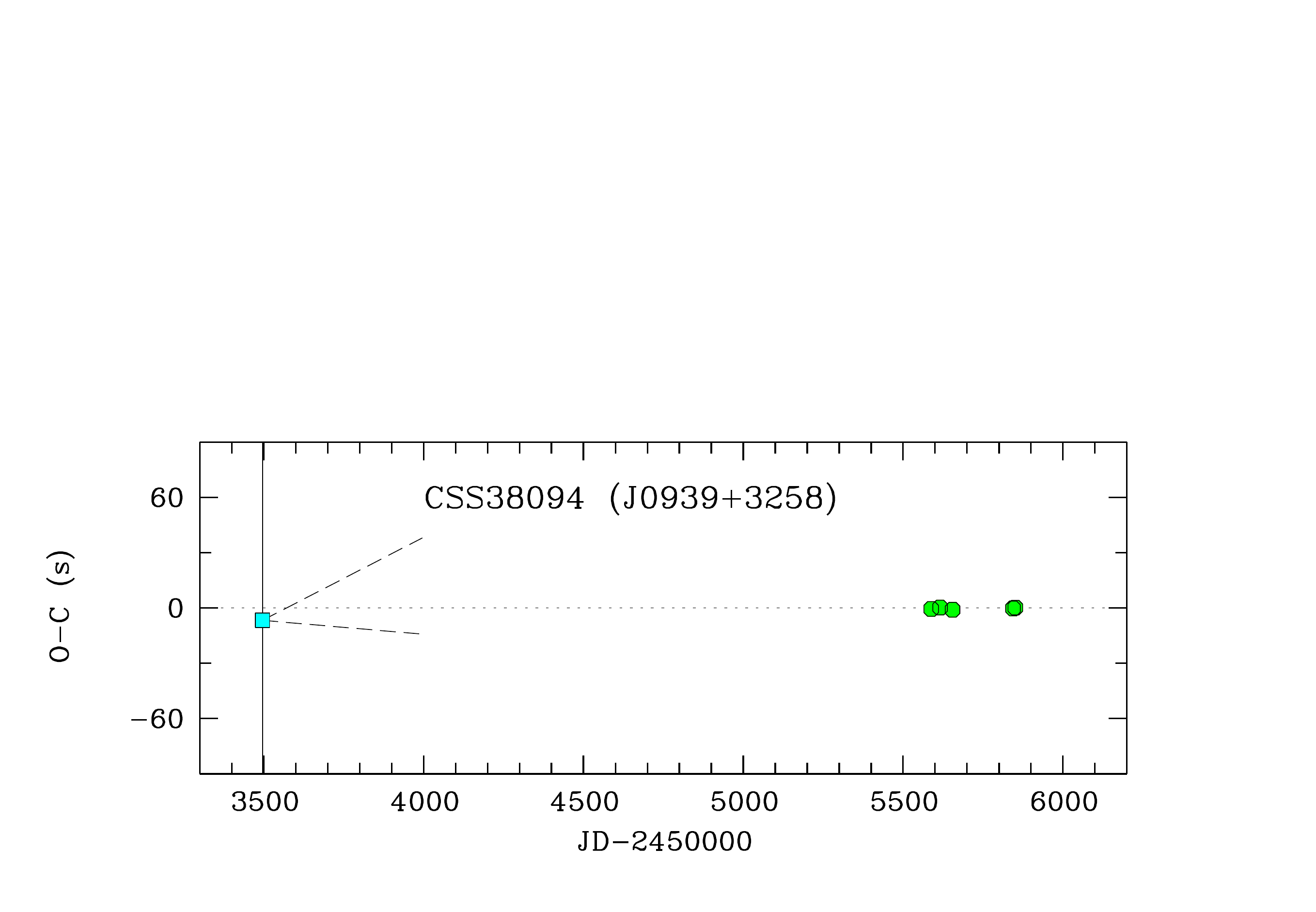}
\includegraphics[bb=260 79 465 700,height=60mm,width=20mm,angle=-90,clip]{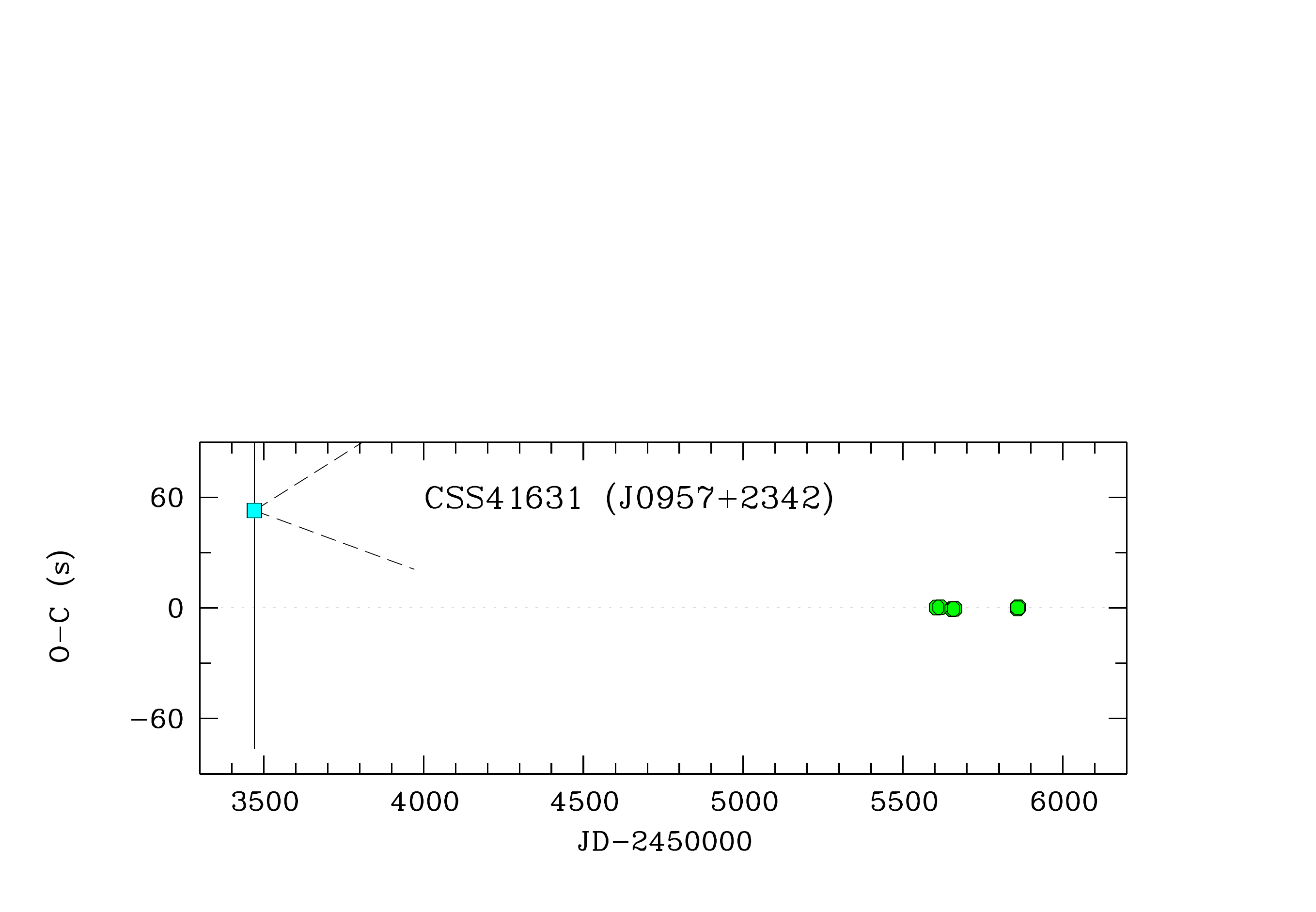}

\includegraphics[bb=260 28 465 700,height=64mm,width=20mm,angle=-90,clip]{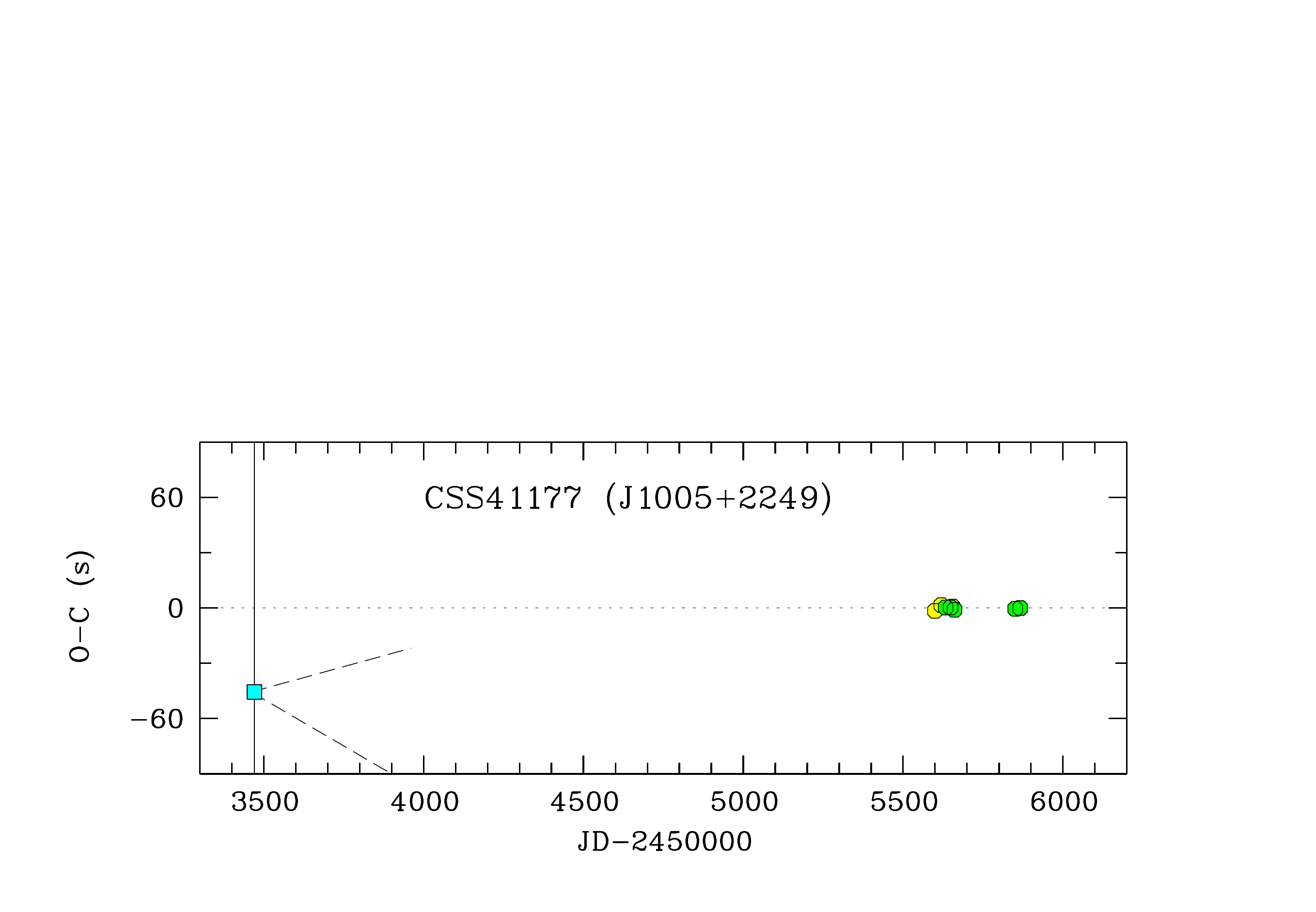}
\includegraphics[bb=260 79 465 700,height=60mm,width=20mm,angle=-90,clip]{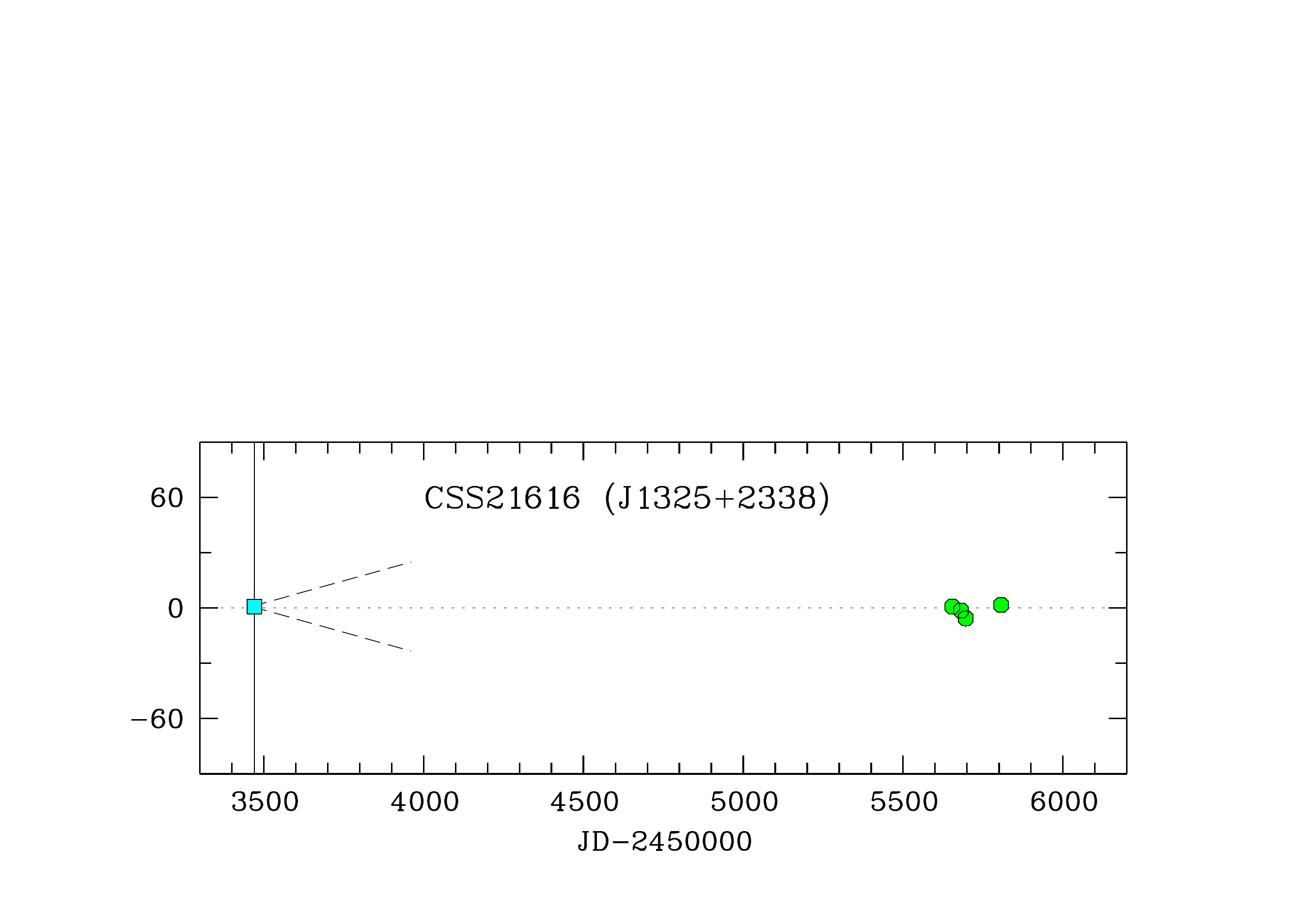}
\includegraphics[bb=260 79 465 700,height=60mm,width=20mm,angle=-90,clip]{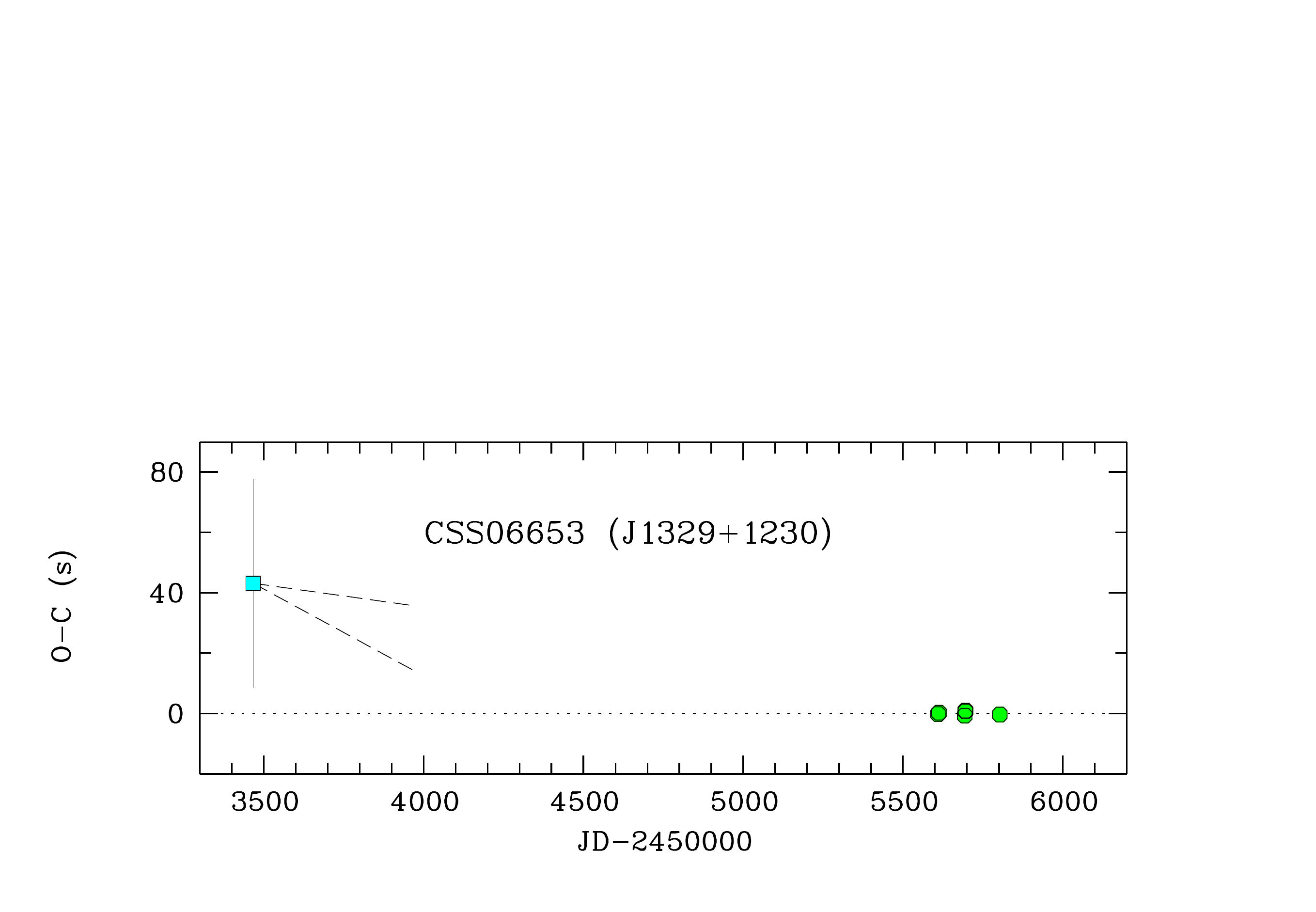}

\includegraphics[bb=260 28 510 700,height=64mm,width=27mm,angle=-90,clip]{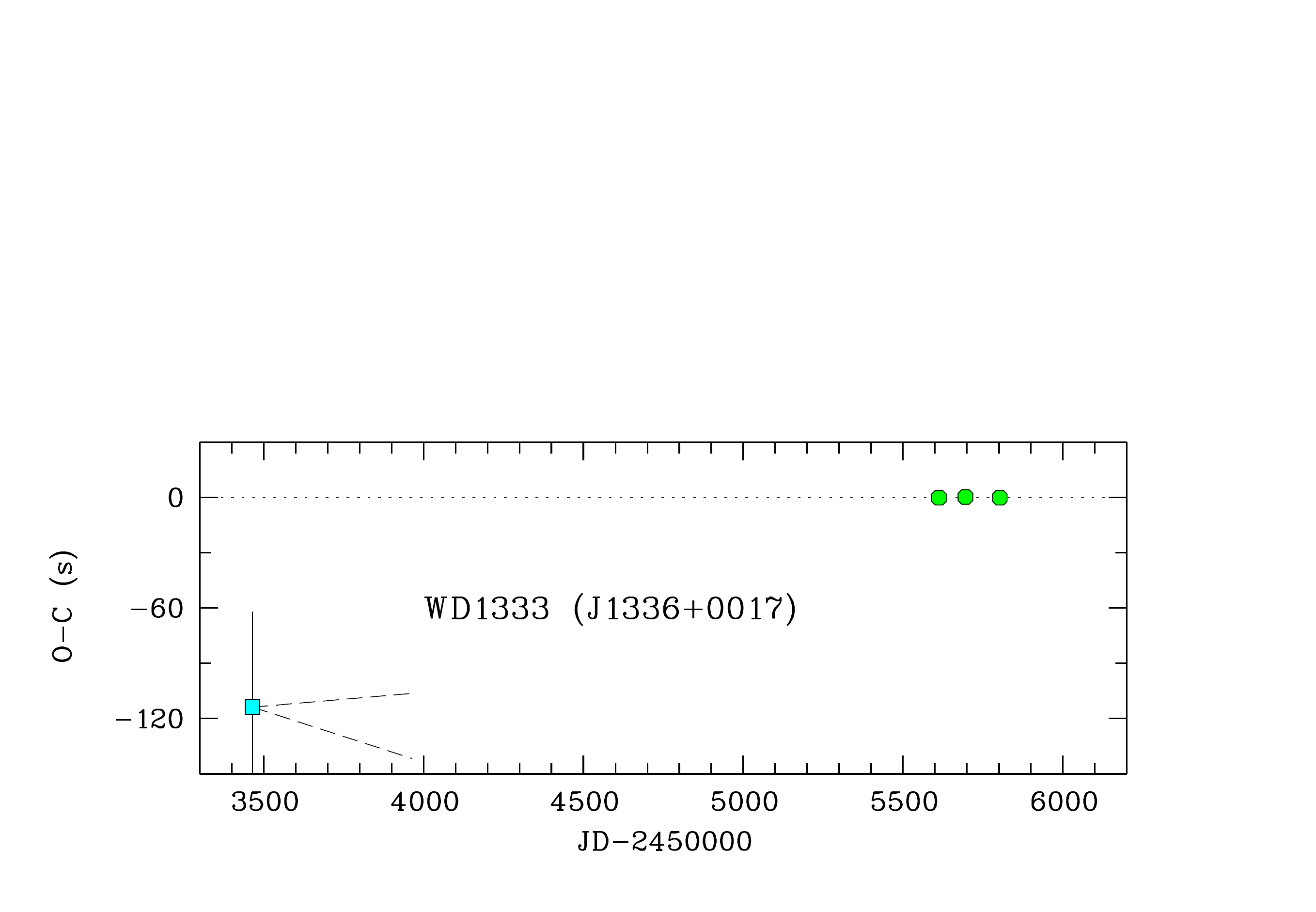}
\includegraphics[bb=260 79 510 700,height=60mm,width=27mm,angle=-90,clip]{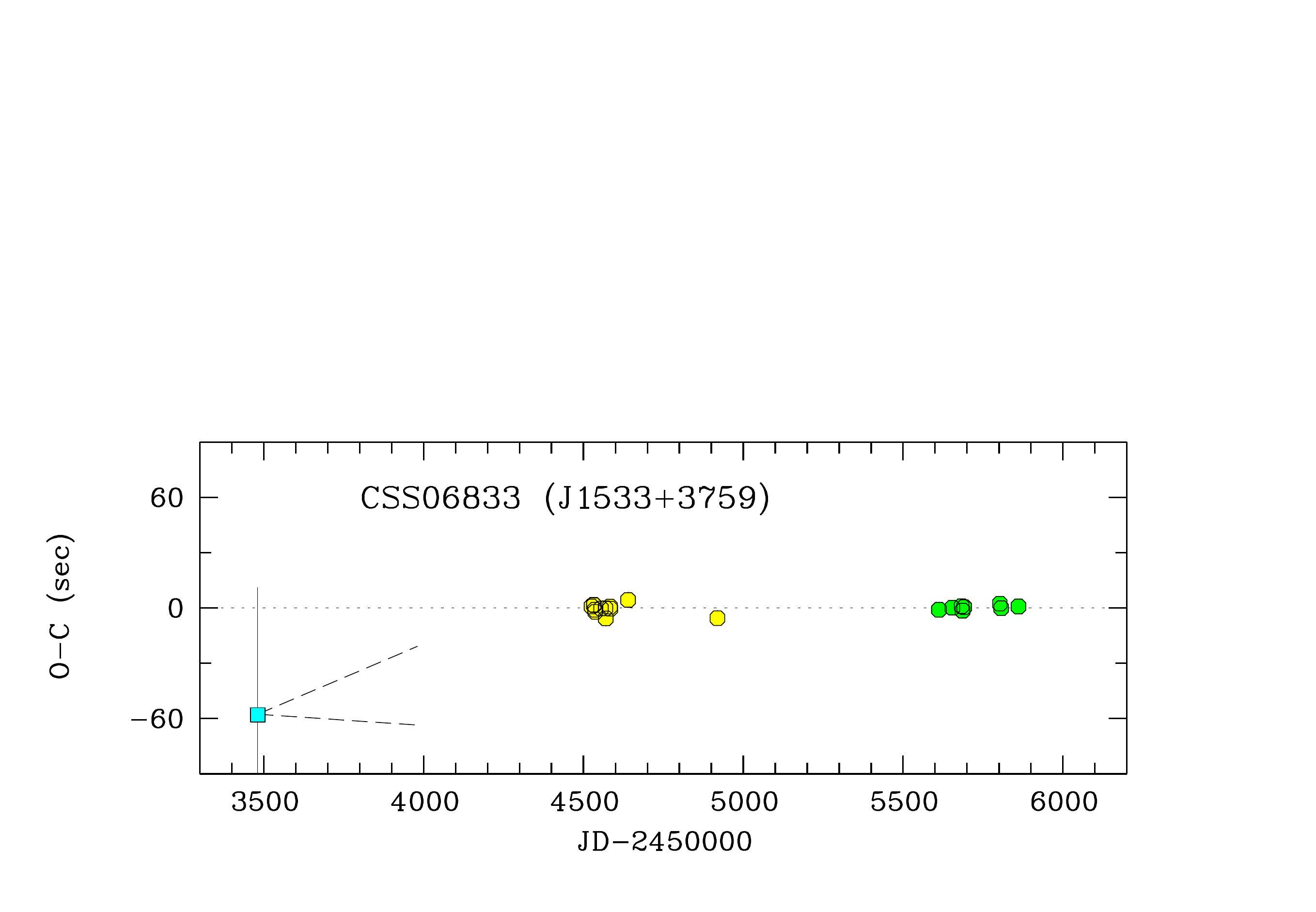}
\includegraphics[bb=260 79 510 700,height=60mm,width=27mm,angle=-90,clip]{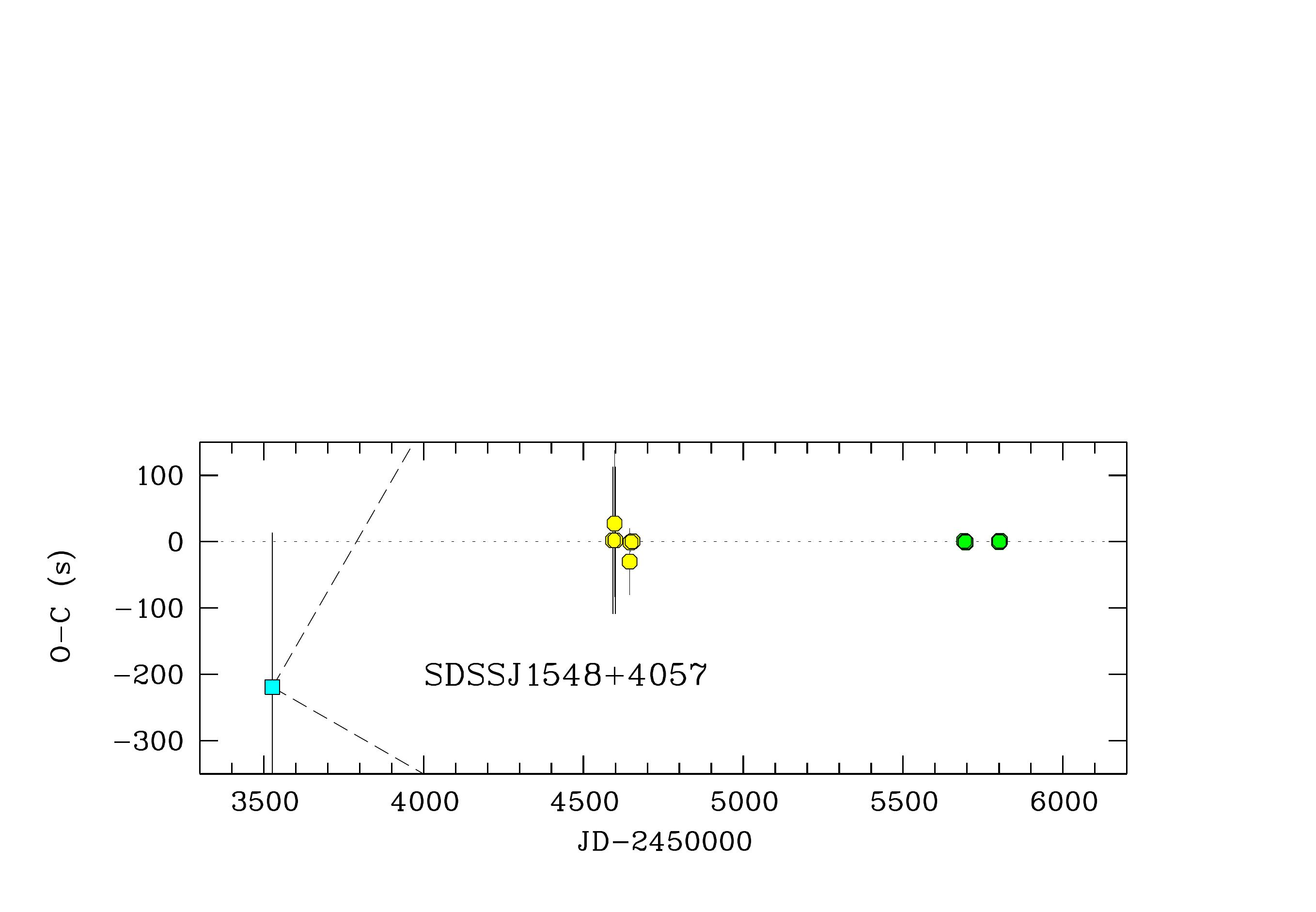}
\caption[chart]{$O-C$ residuals for the new mid-eclipse times of
  Table~1 (green dots) and other published times for
  SDSS\,J0303+0054, CSS\,41177, CSS06833, and SDSS\,J1548 +4057
  (yellow dots) relative to the new ephemerides of Table~2, columns 5
  and 6. The cyan-blue square indicates the Drake et al. (2010) epoch
  (see text).}
\label{fig:oc}
\end{figure*}

So far, we find no evidence for a long-term period variation for any
of the sources. For our own data, this is not surprising, because we
covered only a single observing season.  In the case of
SDSS\,J0303+0054, CSS\,06833, and SDSS\,J1548+4057, though, the data
cover 5.1, 3.6, and 3.3 years, respectively, and are still consistent
with linear ephemerides. In principle, the comparison of the
independent ephemerides of \citet{drakeetal10} and this work could
provide information on a period change between 2005 and 2011 by either
a departure of the Drake et al. epoch from the respective new
ephemeris or by a difference of the periods measured at the two
epochs.  The observational situation is illustrated in
Fig.~\ref{fig:oc}, where in each panel zero $O-C$ represents our new
ephemeris and the dashed lines attached to the Drake et al. points
represent the $\pm1$-$\sigma$ uncertainty in their ephemeris projected
500 days forward in time.  The ordinate scales are chosen to
incorporate the \citet{drakeetal10} epochs. A possible period change
is is taking place in CSS06653, where the epoch and period of
\citet{drakeetal10} combine to a 2-$\sigma$ result of an increasing
period.  The case of WD1333 is also peculiar and calls for more
observations. In summary, period changes that give rise to $O-C$
variations with excursions of several tens of seconds cannot be
excluded for any of the sources. The presence or absence of $O-C$
variations caused by planet-sized or brown-dwarf companions orbiting
the binaries of the present sample remains an open question.

\begin{table*}[t]
\begin{flushleft}
\caption{Ephemerides for the binaries of Table~1, with the epoch quoted as BJD in the terrestrial time system.}
\begin{tabular}{llllllc}
\hline \\[-1ex]
CSS name   & \hspace{6mm}SDSS name & \multicolumn{2}{c}{Drake et al. (2010)} &  \multicolumn{2}{c}{This work} & \hspace{-1mm}$\chi^2$/d.o.f.  \\
& &\hspace{5mm}BJD(TT)   & \hspace{7mm}$P$    &\hspace{5mm}BJD(TT)  &\hspace{7mm}$P$ & \\
& &\hspace{5mm}2400000+  & \hspace{4mm}(days) &\hspace{5mm}2400000+ &\hspace{4mm}(days) & \\[1ex]
\hline \\[-1ex]
            & J030308.35+005444.1 & &                                & 53991.617287(2)  & 0.1344376696(4) & 6.5\,/\,8\\
CSS\,40190  & J083845.86+191416.5 & 53469.72043(130) & 0.13011225(40)& 55569.862961(7)  & 0.130112320(5) & 3.2\,/\,3 \\
CSS\,03170  & J085746.18+034255.3 & 53464.72053(90)  & 0.06509654(3) & 55528.866655(11) & 0.065096539(4) & 5.2\,/\,5 \\
CSS\,080502 & J090812.03+060421.2 & 53466.83093(140) & 0.1494385(25) & 55569.876012(3)  & 0.149438072(3)& 8.2\,/\,9\\
CSS\,38094  & J093947.95+325807.3 & 53495.95413(330) & 0.3309896(2)  & 55587.808823(10) & 0.330989655(21) & 0.4\,/\,3\\
CSS\,41631  & J095719.24+234240.7 & 53470.76412(150) & 0.15087065(15)& 55604.830124(6)  & 0.150870740(6) & 1.3\,/\,4\\
CSS\,41177  & J100559.10+224932.2 & 53470.70403(170) & 0.1160154(1)  & 55619.426445(6)  & 0.116015436(6)   & 3.9\,/\,6\\
CSS\,21616  & J132518.18+233808.0 & 53470.80433(190) & 0.1949588(5)  & 55653.954186(9)  & 0.194958909(42)  & 2.2\,/\,1\\
CSS\,06653  & J132925.21+123025.4 & 53466.81752(40)  & 0.08096622(2) & 55609.022166(4)  & 0.080966254(4)  & 4.9\,/\,4\\
WD1333+005  & J133616.05+001732.6 & 53464.89123(60)  & 0.12195874(5) & 55611.976667(9)  & 0.121958769(11) & 0.2\,/\,1\\
CSS\,06833  & J153349.44+375928.0 & 53480.92373(80)  & 0.16177052(8) & 55611.926580(3)  & 0.1617704531(9) & 16.9\,/\,17\\
SDSSJ1548   & J154846.00+405728.7 & 53526.78503(270) & 0.1855162(15) & 54592.572944(56) & 0.185515296(9)  & 4.6\,/\,10 \\[1ex]
\hline
\end{tabular}
\end{flushleft}
\end{table*}


\section{Discussion}

In recent years, several PCEBs have been found (or suspected) to host
circum-binary substellar objects. The host compact binary stars are
two pulsars, PSR\,1257 and PSR\,B1620, and a small number of post-CE
binaries with either an sdB star or a white-dwarf as primary, both of
the detached and the semi-detached variety. The best cases apart from
the pulsars are probably the detached systems HW Vir
\citep{leeetal09}, NN~Ser \citep{beuermannetal10}, HS0705+67
\citep[][Paper II]{qianetal09}, and the cataclysmic variable DP Leo
\citep{qianetal10,beuermannetal11a}.  The origin of the suggested
companions is uncertain. Either they are of primordial origin and
survived the CE evolution of the host binary or they formed as
second-generation object \citep{perets10} from the expelled envelope
of the compact object \citep[see, e.g., the discussion
  in][]{beuermannetal10}. Any theory of such a scenario will require
information on the frequency of incidence of circum-binary
planets. Obtaining this information requires that a larger number of
binaries are searched for the presence of companions.

The third-body hypothesis lingered in the background for decades
because of the possibility of inducing real or apparent orbital period
variations by other mechanisms.  In detached binaries, these include
spin-orbit coupling induced by variations in the internal constitution
of the secondary star \citep{applegate92} and apsidal motion of an
eccentric binary orbit \citep{todoran72}. The former is generally
thought to be too weak to produce the observed amplitudes
\citep{brinkworthetal06,chen09,potteretal11}, but this may not be the
last word \citep{wittenmyeretal11}. Apsidal motion can be excluded if
the expected shift of the secondary eclipse is found to be absent
\citep[e.g.][]{beuermannetal10,beuermannetal11b}. The high frequency
of exoplanets around normal stars has resulted in a re-consideration
of the third-body hypothesis, starting with a series of papers by the
group of \citet[][and references therein]{qianetal09,qianetal10}.
However, the interpretation of the results is still controversial and
it is not clear whether the same explanation will apply to all PCEBs
that show eclipse time variations.

Here we studied twelve eclipsing PCEBs, mostly identified by
\citet{drakeetal10}, for apparent period variations, which could
indicate the presence of a third body. None were found so far,
primarily because the \citet{drakeetal10} ephemerides lack sufficient
accuracy.  Our substantially more accurate results provide the basis,
however, for a future detection of companions orbiting these
binaries. The sample studied here is a mixed bag that contains systems
with a white dwarf primary, the double degenerate CSS\,41177, and
CSS\,06883 with an sdB primary. The first group includes systems with
a He white dwarf and a CO white dwarf as shown by the masses
derived from SDSS spectra \citep{rebassaetal11}\footnote{see also
  http://astron.dfa.uv.cl/}. Although members of all subgroups went
through a CE phase, their progenitors and evolutionary histories
differ \citep[e.g.][]{zorotovicetal11} and the incidence of
circum-binary planets may differ, too. Elucidating these
connections will require substantially more observational and
theoretical work.


\begin{acknowledgements}
We thank M. Schreiber for discussions and information on the white
dwarf masses in PCEBs. This work is based in part on data obtained
with the MOnitoring NEtwork of Telescopes (MONET), funded by the
Alfried Krupp von Bohlen und Halbach Foundation, Essen, and operated
by the Georg-August-Universit\"at G\"ottingen, the McDonald
Observatory of the University of Texas at Austin, and the South
African Astronomical Observatory.  The "Astronomie \& Internet"
program of the Foundation and the MONET consortium provides a major
part of the observation time to astronomical projects in high schools
worldwide.
\end{acknowledgements}

\bibliographystyle{aa}

\end{document}